# Exploring Cybersecurity Issues in 5G Enabled Electric Vehicle Charging Station with Deep Learning

Manoj Basnet [1], M. Hasan Ali [2]

[1,2] The Department of Electrical and Computer Engineering, University of Memphis, Memphis, TN, USA
[*]mbasnet1@memphis.edu

**Abstract:** The surging usage of electric vehicles (EVs) demand the robust deployment of trustworthy electric vehicle charging station (EVCS) with millisecond range latency and massive machine to machine communications where 5G could act. However, 5G suffers from inherent protocols, hardware, and software vulnerabilities that seriously threaten the communicating entities' cyber-physical security. To overcome these limitations in the EVCS system, this paper analyses the impact of False Data Injection (FDI) and Distributed Denial of Services (DDoS) attacks on the operation of EVCS. This work is an extension of our previously published conference paper about the EVCS. As new features, this paper simulates the FDI attack and the syn flood DDoS attacks on 5G enabled remote Supervisory Control and Data Acquisition (SCADA) system that controls the solar photovoltaics (PV) controller, Battery Energy Storage (BES) controller, and EV controller of the EVCS. The attacks make the EVCS system oscillate or shift the DC operating point. The frequency of oscillation, its damping, and the system's resiliency are found to be related to the attacks' intensity and target controller. Finally, we propose the novel stacked Long Short-Term Memory (LSTM) based intrusion detection systems (IDS) solely based on the electrical fingerprint. This model can detect the stealthy cyberattacks that bypass the cyber layer and go unnoticed in the monitoring system with nearly 100% detection accuracy.

## 1. Introduction

The intelligent and extensive deployment of electric vehicle supply equipment (EVSE) coerces heterogeneous stakeholders and customers to coordinate and communicate. The heterogeneous stakeholders mainly refer to i) the intelligent transportation system (ITS)/vehicle to everything (V2X) infrastructures such as roadside sensors, connected automated vehicles (CAV), electric vehicles (EVs), ii) Electric grid infrastructures such as utility, generation, transmission, distribution, sensors, protection and relays so on, and iii) financial institution such as credit card companies for the management of transactions [1]. The extent of administrative privilege for coordinating these eccentric stakeholders to/from the EVSE is still a conflict of interest due to the lack of clearly developed standards for proper interoperability and a fully matured trustworthy environment [2]. Therefore, the EVSE needs robust, secure, and reliable communication with its stakeholders and customers. In such a scenario, the communication between these multiple nodes may need stringent requirements in terms of latency, bandwidth, and the number of connections. 5G must be the ideal communication tech for fulfilling those requirements. As shown in Fig.1, The most updated deployment scenarios of 5G till now are Industrial internet of things (IIoT) and ultra-reliable low-latency communication (URLLC), extended mobile broadband (eMBB), massive machine-type communication (mMTC), with additions of ITS/V2X, Integrated access and backhaul (IAB) and New Radio based access to unlicensed spectrum (NR-U) [3]–[7].

The general system architecture of EVSE includes the power delivery modules, communication and control modules, sensing and protection modules, and interfaces to users [8]. The power delivery module deals with the unidirectional/bidirectional flow of electric power to/from EV/grid such as a battery, power supply, power regulator.

The communication and control modules extend the capability to communicate with diverse stakeholders such as EV users, operators, utilities, credit card companies, transportation. The key enablers for wireless technology may be 5G/6G, Wi-fi 6, Bluetooth, ZigBee [9]. The sensing and protection modules ensure the good health and safety status of electrical components in the EVSE. These open communications designed to provide robust, smart, and accurate control push air-gapped EVSE physical systems to the edge of cyber-physical vulnerabilities. Some of the vulnerabilities come with the deficiency in communication protocols such as authentication, authorization, and access control, some with the inherent component vulnerabilities. Furthermore, there is always a risk from insider threats and socially engineered advanced persistent threats from the notorious hackers[10].

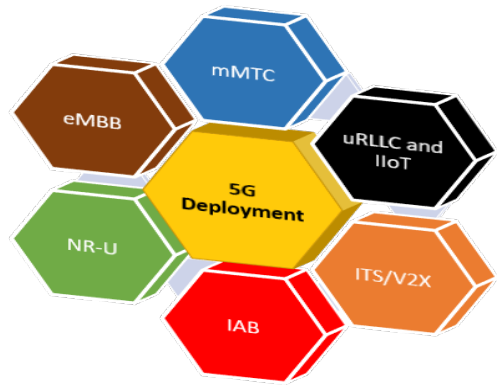

**Fig. 1**. Deployment scenarios of 5G

Cyberattacks' motivation in EVSE ranges from a prank, electricity theft, identity theft to vicious ransomware and malware that could infect the entire EVSE network [11]. The infected EVSE can spread malware to other EVSE via charging EVs that get infected [12]. The transition and transformation of the attack vector from the communication/cyber layer to the physical infrastructure layer are the intricate metrics that should be analyzed in terms of the aftermath in the real physical entities such as power, current, voltage, and state of charge (SOC). The work in [13] enlists the vulnerability analysis and risk assessment of the EVSE with details of the potential attack scenario such as a DoS, man in the middle (MITM), FDI. The scope of the paper will mostly cover DDoS and FDI. One of the DoS attacks, e.g., SYN-flood, originates from the inability of transport control protocol (TCP) 's three-way handshakes to correctly identify legitimate requests from the client's nodes and responding each of them. Therefore all of the DoS attacks target the network/source availability by processing the illegitimate requests assuming it as the legitimate users causing the congestion [14]. Furthermore hence, it is imperative to develop a system that could efficiently identify DoS attacks.

The proactive strategy for accurate detection and classification of the EVSE network attacks on time is known as an intrusion detection system [15]. It helps the operator take protective and preventive measures once the attack is identified. There are three types of IDS derived from their implementation locations, namely: Host-based (HIDS), network-based (NIDS), and hybrid IDS [11]. Three basic intrusion detection techniques have been widely deployed in state-of-the-art applications, namely: Signature-based detection (SD), Anomaly-based detection (AD), and Stateful protocol analysis (SPA) [11] [16].

Researchers listed and characterized exploitable backdoors of the EV charging infrastructure; however, they lack the impact analysis of attacks and detection and mitigation strategies [17]. Authors [18] presented a system approach to list the interactions between various cyber-physical components inside the smart EVSE and few approaches to improve its cyber-physical security. This research work also lacks the analysis of the impact on EV charging and any proactive detection techniques. The research work [19] introduced the concept of cyber insurance that transfers the risk of paying a high price from user to third party in offline EV charging. Nevertheless, the cyber insurance model lacks security analysis. Likewise, cyber threats targeting different players in an EVSE are presented, lacking impact analysis and mitigation techniques [20]. Apparently, most of the research [17]-[20] in EVSE cybersecurity is limited to vulnerability analysis and risk assessment; and unable to explain the exact quantifiable effects on the physical system from the cyber-initiated attack. The literature also lacks the proactive attack detection strategy such as IDS [13] and post-attack specifications to deal with the attack in standalone EVSE or fleet.

The automation of EVSE operation and management requires remote centralized control like SCADA to communicate with numerous field devices with the least possible delay. Since 5G is the proven cellular technology with less than 1 ms latency capable of million of machine type communication [21], it can be the candidate technology for EVSE communication too.

Based on the above background, this paper steps ahead to propose and simulate the 5G enabled PV-powered standalone EVSE architecture controlled by a remote centralized controller/SCADA. The centralized controller can control the entire EVSE station to manage all the controllers, namely a PV controller, BES controller, and the EV controller. We developed the stealthy FDI and the SYN-flood DDoS attack strategy, ran the proposed architecture in a simulated environment, and presented an impact analysis of attack penetration on EVSE. Moreover, we proposed the novel deep learning-based local IDS to detect the attack at the infrastructure layer/physical layer rather than the network/cyber layer. The local IDS' motivation is that a stealthy attack may pass through the IT system and disrupt and dismantle the physical infrastructure of EVSE. Therefore, the proposed model could detect the bypassed attack by using the fingerprint of electrical parameters. So, the proposed model would make the EVSE smarter by making it self-reliant against cyberattacks.

The proposed IDS implements the variant of recurrent neural network (RNN) called stacked or deep LSTM specifically designed to deal with sequential learning. The motivation for using LSTM came from the sequential nature of electrical and control signals and its superior classification performance in our past work [11]. Unlike current work, our past work solely depended on network packet fingerprint to train the deep learning models to detect the cyberattacks at the network layer. It can not detect the bypassed attack at the infrastructure layer.

This research work extensively used the NetSim standard version 12.2 for network simulation that acts as a bidirectional communication link of the remote SCADA managing the PV, BES, and EV controllers at the EVSE. To implement a testbed for the PV-powered standalone EVSE architecture with BES and EV charging and to assess the impact of FDI and DDoS attacks, MATLAB 2020b and Simulink were used. Also, Python 3.7 was used for stacked LSTM based IDS implementation to detect and classify the attacks. This work is an extension to our previously published conference paper about the EVCS [11]. However, the main contributions of this paper are summarized as follows:

1) Development of a simulation model of EVSE system including a standalone PV, BES, EV charger, and associated controllers.
2) Design and simulation of FDI and DDoS attacks strategy through 5G enabled remote SCADA.
3) Impact assessment in terms of electrical parameters and fingerprint acquisition.
4) Implementation and testing of the novel stacked LSTM IDS based on electrical fingerprints to detect and classify the attacks at different controllers in the physical layer.

The organization of the paper is as follows: Section 2 describes the proposed EVSE architecture. Section 3 gives an overview of cybersecurity issues in EVSE. Section 4 describes cyberattack modeling. Section 5 presents the proposed IDS methodology. Section 6 presents the simulation procedure. Section 7 presents the simulation results and discussion. Finally, Section 8 concludes the paper.

## 2. Proposed EVSE Architecture
### 2.1. EVSE Architecture

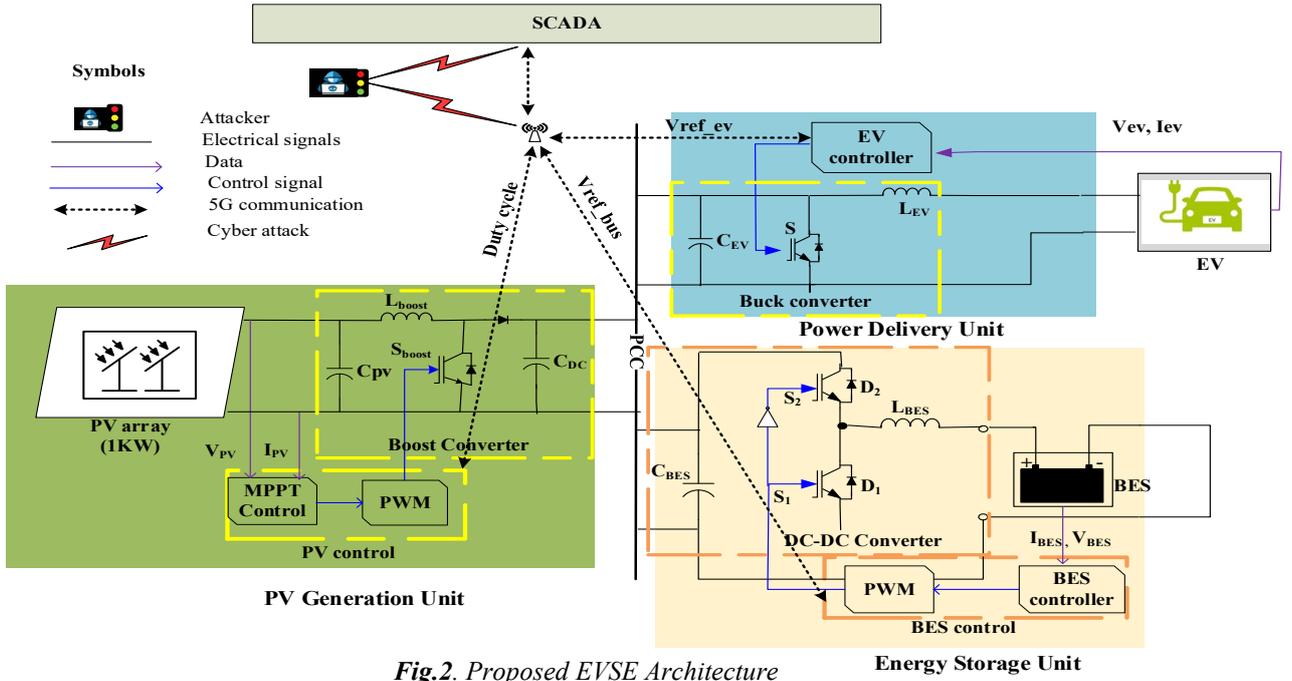

*Fig.2. Proposed EVSE Architecture*

Our proposed EVSE is a standalone, PV-powered, and off-the-grid system. It has three electrical units: PV Generation Unit (PGU), Energy Storage Unit (ESU), and Power Delivery Unit (PDU), as shown in Fig. 2. The PGU consist of a PV array, a boost converter, and a control circuitry. The PV arrays deliver 1.065 kW at maximum power point (MPP) with the corresponding voltage of 36.75 V and current of 29 A at the constant irradiance of 1000 W/m² and constant temperature of 25 °C. The boost converter boosts the PV voltage ($V_{PV}$) to the DC link bus bar voltage of $V_{ref\_bus}$. The ESU consists of BES, a DC-DC converter, and a control circuitry. The BES has a nominal voltage of 48 V and a nominal discharge current of 43.47 A with a 100 Ah rated capacity. The DC-DC converter charges the BES in buck mode while there is a surplus generation and discharges to the bus bar in boost mode while the PV can't meet the EV demand. The control circuitry continuously senses the battery current ($i_B$) and $V_{ref\_bus}$. Thus generates the pulses to drive the bidirectional DC-DC converter (BDC). PDU could be an offboard in the EVSE or onboard integrated with the EV. Either way, the functionalities remain the same. PDU has a buck converter and an EV controller. The buck converter steps down the voltage from the point of common coupling (PCC) as per the requirement of EV, i.e., $V_{ref\_ev}$. The EV controller continuously monitors the status of EV battery voltage ($V_{ev}$) and battery current ($I_{ev}$) and generates the pulse to adjust the switching of Buck Converter. Here, all three control units, namely, PV control, BES control, and EV control, can be assessed/overridden by remote operators at SCADA, EVSE, or EV owner through apps or web via robust 5G communication infrastructures.

Control circuitry: As shown in Fig. 2., there are three controllers: the PV control, BES control, and EV control. These are explained in detail below.

PV control: The MPPT control continuously reads the $V_{PV}$ and $I_{PV}$ signals from the PV output. The Perturb and observe Maximum Power Point Tracking (P&O MPPT) algorithm tracks the maximum power points and corresponding $V_{PV}$ and $I_{PV}$ and adjust the duty cycle accordingly. The pulse width modulation (PWM) circuitry block will generate the $S_{boost}$ signal from the duty cycle. The details of P&O MPPT algorithm can be found in [22]. The duty cycle of MPPT can be adjusted and reinitialized manually by a human operator in the case of malfunction, disaster, and emergency. This operator at SCADA can remotely monitor and control the PV controller via the 5G communication.

BES control: The BES controller continuously monitors the $I_{BES}$, $V_{bus}$, and $V_{ref\_bus}$ from the system. The slower outer loop in Fig.3a. controls the bus voltage with the help of a Proportional-Integral (PI) controller driven by the error $V_{ref\_bus} - V_{bus}$. It generates the reference signal $I_{bes\_ref}$ for inner current control loop which is ten times faster than the outer loop as in Fig. 3b.

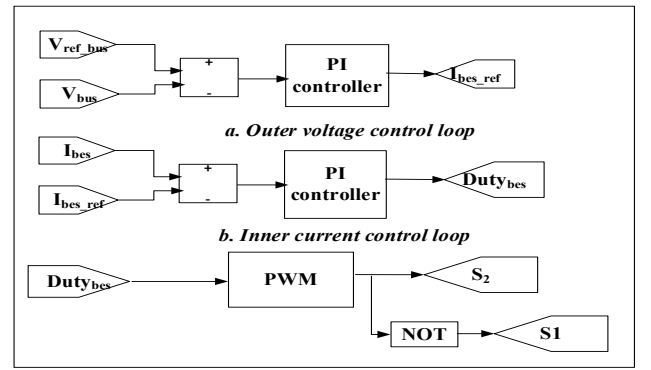

*Fig.3. BES control a) Outer voltage control b) Inner current control.*

The PI controller tries to track the $I_{bes\_ref}$ by minimizing the error between reference current and measured current and generates the duty signal $Duty_{bes}$. This $Duty_{bes}$ drives the PWM to create complementary pulses $S_1$ and $S_2$ that trigger the switching of boost and a buck converter, respectively, in the bidirectional DC-DC converter. The details of this cascaded PI control strategy can be found in [23].

Through the 5G, the SCADA operator at the remote station could wirelessly monitor and control the BES controller at EVSE and can set the $V_{ref\_bus}$ as well as other PI controller settings.

<u>EV control</u>: The EV controller continuously monitors the battery's voltage ($V_{Bev}$) and current ($I_{Bev}$) of the PEV. This control block might be offboard or onboard the EV. The reference battery voltage ($V_{ref\_ev}$) can be set by EVSE owner or SCADA operator if it is offboard or can be set by EV owner for dynamic charging or hardcoded by original equipment manufacturer (OEM) in CAN (Control Area Network) bus if it is onboard. These communications may take place through the 5G. The same cascaded outer voltage and inner current control strategy, as in BES control, are implemented to generate the Dutybev, which controls the buck converter to regulate the EV charging, as shown in Fig.4 below.

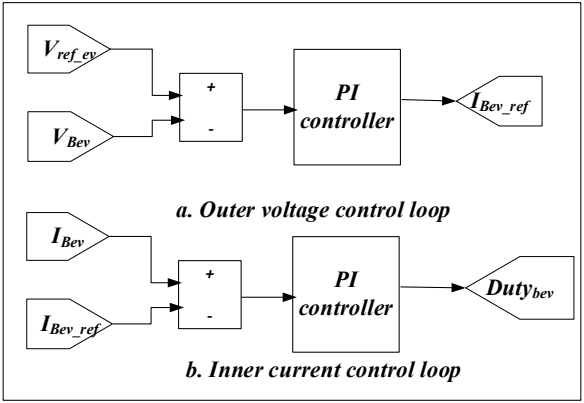

**Fig.4.** *EV controller a) Outer voltage control b) Inner current control*

### 2.2. System formulation and component Modelling
<u>PV array</u>:

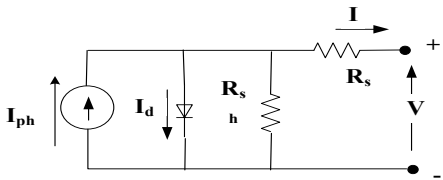

**Fig.5.** *One diode equivalent circuit of PV module.*

The mathematical representation of the one-diode equivalent circuit model of a PV system is given by transcendental equation [24] as in eq. (1).

$$I = I_{ph} - I_0 \cdot \left\{ e^{\frac{q}{N_{cs}\cdot k\cdot T\cdot \gamma}\cdot(V+I\cdot R_S)} - 1 \right\} - \frac{V+I\cdot R_S}{R_{Sh}} \quad (1)$$

Where I and V are the current and voltage of the PV module, respectively. $N_{cs}$, $k$, $T$, and $q$ denote the number of cells in series, Boltzmann constant, cell temperature, and elementary charge, respectively. The free model parameters are photocurrent $I_{ph}$, diode saturation current $I_0$, series resistance $R_S$, shunt resistance $R_{Sh}$, and the diode ideality factor $\gamma$.

The photocurrent depends on both irradiation $G$ and temperature $T$ as in eq. (2) [25].

$$I_{ph}(G,T) = \frac{G}{G_{ref}} \cdot [I_{phref} + \mu_{I_{SC}} \cdot (T - T_{ref})] \quad (2)$$

Where $G_{ref}$, $I_{phref}$, and $T_{ref}$ are the irradiance, photocurrent, temperature at some arbitrarily chosen reference conditions with $\mu_{I_{SC}}$ representing the temperature coefficient of $I_{SC}$.

<u>Boost converter</u>: The boosting of voltage in the boost converter depends on the duty ratio $D_b$ as in eq. (3) [26]. Also, the boosting parameters L and C can be further calculated as eq. (4) and (5), respectively.

$$V_{DC} = \frac{1}{1-D_b}V_{PV} \quad (3)$$

$$L_{boost} = \frac{V_{DC}\cdot D_b}{\Delta I_L f} \quad (4)$$

$$C_{DC} = \frac{V_{PV}D_b}{R_0 \Delta V_{PV} f} \quad (5)$$

Where $V_{PV}$, $V_{DC}$, $\Delta I_L$, $V_{PV}$, $R_0$ and $f$ are input voltage from PV, an output voltage of converter, inductor ripple current, capacitor ripple voltage, the output impedance of boost converter, and switching frequency, respectively, in Fig. 6a.

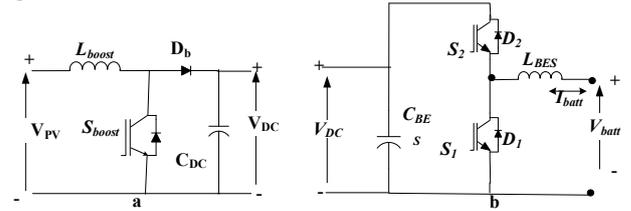

**Fig.6.** *a) Boost converter b) Bidirectional DC-DC converter.*

<u>Bidirectional DC-DC converter</u>: The inductance and capacitance for buck mode of BDC, as in Fig. 6b, while charging is given by eqs. 6 and 7.

$$L_{buck} = \frac{(V_{DC}-V_{batt})\cdot D_{buck}}{\Delta I_L f} \quad (6)$$

$$C_{buck} = \frac{(1-D_{buck})V_{batt}}{8L_{buck}\Delta V_{batt} f^2} \quad (7)$$

Similarly, for discharging mode, i.e., the BES charging EV controller should act as Boost converter and follow the eqs. 4 and 5.

### 2.3. 5G Architecture
The virtualization of the core network (CN) in 5G, unlike its predecessor, adds higher flexibility and portability for the network resource management that delineates the control and user plane separation (CUSP) with the help of Software-defined network (SDN) and Network function virtualization (NFV) [27]. SDN and NFV complement each other for simpler network control and management, better elasticity, and eliminating vendor-specific solutions [28]. Besides CN, SDN, and NFV, management and network orchestrator (MANO), multi-access edge computing (MEC) are key enablers of 5G. The physical infrastructures layer embodies storage, computing, and networking infrastructures. The virtualized infrastructure has a 5G radio access network (RAN), IAB, 5G core network functions (5G CN NF), MEC, and data network (DN) with MANO for additional network slicing. The ENISA has added security architecture processes on its December release because these are the prime stakeholders towards 5G security. The added processes are mobile network operators (MNO), assurance, and vendors. The functional diagram of security architecture and the in-depth information of individual components could be found here [28].

## 3. Cybersecurity issues in EVSE

The communication system is the nerves of the EVSE that facilitates various operations such as EV scheduling, slot allotment, authentication and authorization, charging session control, grid integration, and on and on [11]. Therefore, once the communication is compromised, the entire EVSE is at the risk of disruption and dismantling. The future grid is envisioned to handle the bidirectional power flow, and blockchain-assisted peer-to-peer energy transactions, and vehicle to everything communication (V2X) [29]. The proper communication technology like 5G should moderate this odd marriage of evolving technology and the traditional grid infrastructure. Once the critical infrastructure is exposed to the open cyber layer through communication links, It is no more secure [19]. The communication vulnerabilities can be exploited to get access to the SCADA or EVSE system.

An attacker may use social engineering such as phishing and/or reverse engineering to get the legitimate SCADA's or EV's credentials. Then, the attacker can impersonate the legitimate SCADA operator or EV owner to breach the system security [11].

### 3.1. 5G enabled EVSE cyber-physical security threats landscape

Threat actors first identify the 5G assets' vulnerabilities and then exploit them by assessing the attack surfaces. Table 1 depicts the CIA triad of 5G assets indicating the assets' risk [28].

Based on the works [23]–[26], any cyber-physical security threats can be mainly classified into four different categories: Nefarious act/Abuse, Eavesdropping/ Hijacking/Interception, Intentional and/or Accidental damages, and Outages. The first two are ill-willed malicious actions generally targeted in cyberspace, while the latter two are threats to physical security. Nefarious activity/ Abuse targets the ICT infrastructures to steal, alter, or destroy the target. Eavesdropping/Hijacking/Interception target unauthorized communication links to listen, seize, or interrupt the services. The intentional/ unintentional damage is intentional/unintentional action that causes damages/harms to the physical infrastructures and persons. Outages are the category that disrupts the availability and quality of service. Threats from nefarious activity/abuse are the most prominent and damaging threats for both 5G and EVSE infrastructures. Some of them are listed below:

**Table 1.** CIA triad of 5G assets

| Asset | C | I | A |
|---|---|---|---|
| MANO |  |  |  |
| Network products |  |  |  |
| Interconnections |  |  |  |
| Services |  |  |  |
| Organizations |  |  |  |
| protocols |  |  | N/A |
| Data |  |  |  |
| Processes |  |  |  |

Red= Very high, yellow= high, green= medium

**Denial of service:** The prime target of DoS is to disrupt 5G/EVSE service availability. DoS can be triggered in many ways, such as botnet/DDoS, flooding of network components/base stations, jamming/interfering with the radio frequency, replay, amplification attacks, etc.

**Malicious code:** The injection of malicious code to the software environment detriments and affects the processes, control actions, and operating conditions of the system. Some examples are viruses, malware, rootkits, worms, trojans, rogueware, ransomware, and SQL and XSS injection attacks.

**Exploitation:** Most of the hardware and software system has glitches or weaknesses. The attacker can exploit vulnerabilities in architecture, design, and configuration of the network and software/hardware such as zero-day exploits, open API, and edge API exploits.

**Abuse:** Since 5G based EVSE is a highly complex, heterogeneous cyber-physical system with poorly developed administrative coordination and control, there are immense potentials of abuse of remote access to the network, authentication/authorization, information leakage, virtualization, and even lawful interception.

**Manipulation:** An insider/outsider attacker can compromise/manipulate hardware equipment, control settings, data, network resources. They might attempt MAC spoofing, memory scraping, side-channel attacks, fake nodes, rouge MEC gateway, UICC format exploitation.

Besides that, there are always imminent threats from compromised vendors, spectrum sensing, data breach, unauthorized activities, identity theft/spoofing, signaling storms/frauds.

## 4. Cyber-Attack Modelling

Our work has the following assumptions: to initiate the DoS attacks; hackers used the spoofed IP of legitimate EVSE. The channel loss of the 5G network is set to be zero. We have used our dataset generated from the proposed system to test our deep learning algorithms.

### 4.1. FDI attack Modelling

A $N_m$-dimensional measurement vector $y$ of any nonlinear EVSE system function $H$ depends on $N_n$-dimensional system state variable vector $x$ with normally distributed $N_m$-dimension measurements error vector $e$ as in eq. 8 below [30].

$$y = Hx + e; \quad e \sim N(0,1) \qquad (8)$$
$$\hat{x} = \arg\min_x J(x) = (y - Hx)^T W(y - Hx) \qquad (9)$$

The overdetermined system (where $N_m > N_n$) may not have an exact solution. Therefore, the attacker estimates the system variable $\hat{x}$ by using any of the lightweight optimization algorithms such as mean square error, least square error, or log likelihood. We are using weighted least square error over the residual function $J(x)$ as in eq. 9. Where W is the weight matrix and defined as $diag\{\sigma_1^{-2}, \sigma_2^{-2}, ..., \sigma_{N_m}^{-2}\}$ and $\sigma_i^2$ is the variance of $i^{th}$ measurement. The $y$ is identified as FDI if it exceeds the predetermined residual threshold (Euclidean norm) [31] $\tau$ as in eq. 10.

$$J(\hat{x}) = (y - H\hat{x})^T W(y - H\hat{x}) > \tau \quad (10)$$
$$y_a = y + a = Hx + a \quad (11)$$

Let $y_a$, in eq. 11, be the measurement vector under the FDI attack vector $a$ having the same dimension as $y_a$. The attacker can access the logged data $y$ and limited state variable $x$ during the reconnaissance phase of the attack.

The attacker can choose the distribution of attack vector $a$ randomly or based on some heuristics. The more sophisticated and stealthy attack can be launched without being caught but the impact may not be enough to disrupt the normal operation. The stealthy false data $\hat{x}_a$ can be estimated using some nonlinear functions $g$ as eq. 12 with the help of eq. 11. Now the stealthy attack's objective is to get the attack vector $a$ that maximizes the error injected into the system without exceeding the detection threshold of $\tau$ which is the constrained optimization problem as in eq. 13.

$$\hat{x}_a = g(y_a) = g(y + a) \quad (12)$$
$$\max_a ||\hat{x}_a - \hat{x}|| \text{ subject to } (y - H\hat{x})^T W(y - H\hat{x}) < \tau \quad (13)$$

Based on the above background, the attacker can launch an FDI attack at three different controllers: PV controller on duty cycle, BES controller on $V_{ref\_bus}$, and EV controller on $V_{ref\_ev}$. The attacker can solely control the duration of the attack and distribution of false data. The eqs. 14-16 represent the FDI attack vector for PV control, BES control, and EV control, respectively. The $PRN$ (0,1,10) stands for a pseudorandom number that fluctuates ten times between the lower bound of 0 and upper bound of 1. The reason for choosing PRN is completely heuristic-based as the duty cycle ranges within this limit. Similarly, the attack injection at both BES and EV follows the Gaussian distribution (G) with respective mean and variance as in eqs. 15 and 16.

$$\widehat{D}_a = D + \Delta D \; ; \; \Delta D \sim PRN(0,1,10) \quad (14)$$
$$\widehat{V}_{ref\_bus_a} = V_{ref\_bus} + \Delta V_{ref\_bus} \; ; \; \Delta V_{ref\_bus} \sim G(48,10) \quad (15)$$
$$\widehat{V}_{ref\_ev_a} = V_{ref\_ev} + \Delta V_{ref\_ev} \; ; \; \Delta V_{ref\_ev} \sim G(24,10) \quad (16)$$

### 4.2. DDoS attack modeling

It is considered that the remote SCADA station continuously monitors all three control stations through the 5G. Once the DDoS launched through the 5G core network, there would be no signal reaching the EVSE. The duration of signal lost depends on the communication delay of the 5G network. The SCADA issues a control signal $\{x_i\}_{i=0}^{N}$ at any timestamp i with N being total numbers of samples. If the communication delay caused by a DDoS attack in 5G network is $N_0$, then the original control signal and delayed signal are presented in eq. 17 and 18, respectively.

$$x_{orig}(n) = x(n) = \{x_i\}_{i=0}^{N} \quad (17)$$
$$x_{del}(n) = x(n - N_0) = \{x_i\}_{i=-N_0}^{N-N_0} \quad (18)$$

In other words, to get the same sample from the original control signal $x_{orig}$, we should add $N_0$ sample time to the current timestep to $x_{del}$ as in eq. 19.

$$x_{del}(n + N_0) = x_{orig}(n) \quad (19)$$

The attack sequence goes like this: a) EVSE working normally i.e., communicating normally with remote SCADA through 5G preattack as in eq 20a. b) Suddenly, the DDoS attack starts at sample time of $n_1 \geq 0$ and lasts up to $n_2 = n_1 + N_0 \leq N$ where $N_0$ is the variable delay that depends on the severity of DDoS attack and comes from NetSim 5G simulation. At this time, the signal is completely lost, i.e., zero as in eq 20b. c) After the attack is gone, the signal should retain the $n_1 + 1$ signal sample as in post-attack of eq 20c because DDoS should not compromise with the signal integrity. The composite control signal reaching to the EVSE controller $x_{EVSE}$ can be expressed as eq. 20.

$$x_{EVSE}(n) = \begin{cases} x_{orig}(n) & \text{if } 0 \leq n < n_1, \text{pre attack (a)} \\ 0 & \text{if } n_1 < n \leq n_2, \text{attack (b)} \\ x_{del}(n) & \text{if } n_2 < n \leq N, \text{post attack (c)} \end{cases} \quad (20)$$

### 4.3. DDoS Attack Launch through 5G

TCP-SYN flood attack is the type of DDoS Attack that exploits the vulnerability of three-way handshake in TCP protocol of machine to machine communication. The attacker sends the TCP connection requests faster than the targeted machine can process, culminating in network saturation [32]. As shown in Fig. 7, The Attack can be summarized as follow: a) Client EVSE sends TCP packet with SYN flag on using 5G network, b) SCADA server on receiving the SYN packet sends back SYN-ACK packet to the client EVSE leaving half-open port for up to TCP connection timeout period. c) EVSE acknowledges the SYN-ACK packet by sending an ACK to the SCADA server, and the communication starts. Before the half-open connections expire, malicious EVSE either impersonating the legitimate EVSE or spoofing the IP sends myriads of SYN requests to create many more half-open connections [33]. The malicious EVSE never receives SYN-ACK in spoofed IP and never sends ACK, coercing the SCADA server to wait forever.

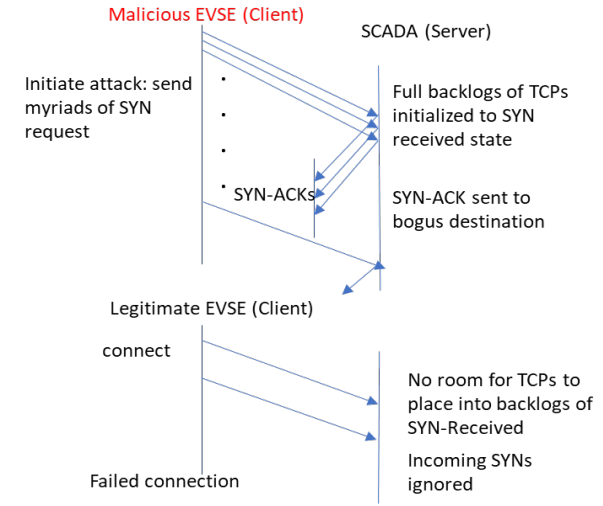

*Fig. 7. SYN-Flood Attack*

In this experimentation, the Attacker or Rouge EVSE node sends TCP-SYN requests every 1000 microseconds.

Upon receiving the SYN request, the SCADA server/Target acknowledges the SYN request by sending the SYN-ACK packet and hold communication open while waiting for the EVSE to acknowledge the open connection. The processing time for SYN-ACK is 2000 microseconds in ethernet physical out delays the arrival of SYN-ACK at the EVSE node. A rouge EVSE node creates another SYN packet exploiting this delay. Over time, it creates an SYN request score exhausting all the available server resources resulting in a successful SYN flood DoS attack as specified in [34].

## 5. Proposed IDS Methodology

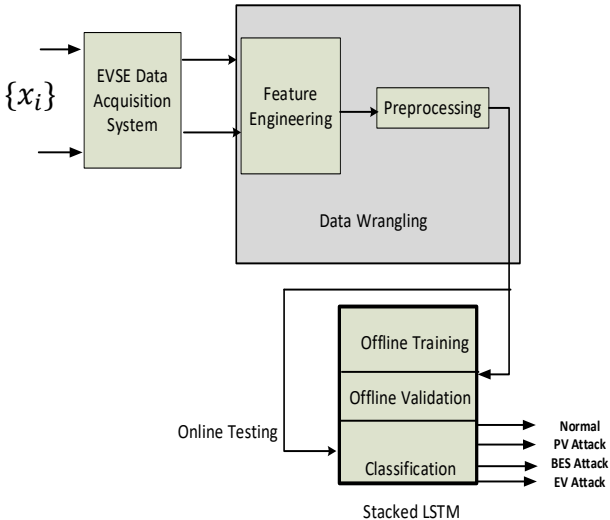

**Fig. 8.** *Proposed IDS at EVSE.*

The proposed IDS is tested and validated with FDI attack on our EVSE architecture. As in Fig.8, the proposed IDS constantly records the logs of electrical and control signals (fingerprint) through the Data Acquisition System. Then the Data Wrangling block chooses, processes and fetches the features into stacked LSTM in the compatible format. First, the output class for each data sample has been assigned. That can be done by appending {'0'}, {' 1'}, {'2'}, {'3'}, {' 4'} at the end of dataset, i.e., 37th column of each class. Now, these strings at the output class have been converted to categorical data for classification purposes. The dataset can be splitted into two mutually exclusive sets: train (70%) and test (30%) class. 20% of training data were further splitted into validation data. Training data in deep learning is used to fit the nonlinear convex curve by using input-output mapping based on forward and backward propagation. Validation data are generally used for hyperparameter and model tuning. Testing data is used to see the generalization of the trained model. Our deep LSTM model is 3 layered with a 10% dropout between each stacked LSTM layer, and it has an output layer with four nodes. The deep LSTM architectures and parameters are shown in table 2 below.

**Table 2.** stacked LSTM architecture

| Layer(type) | Output Shape | Param # |
|---|---|---|
| LSTM_1 | (None,None,64) | 25856 |
| Dropout_1 | (None,None,64) | 0 |
| LSTM_2 | (None,None,64) | 33024 |
| Dropout_2 | (None,None,64) | 0 |
| LSTM_3 | (None,64) | 33024 |
| Dropout_3 | (None,64) | 0 |
| Dense_1 | (None,64) | 260 |

Each stacked hidden layer has 64 LSTM units. The model is compiled using the categorical cross-entropy loss function and Adam optimizer, which are de facto choices for multiclass classification. The model is trained for 20 epochs with a batch size of 1000 samples that took almost 32.91 minutes.

The primary goal of the proposed system is to detect the different classes of bypassed cyberattack on different components of EVSE system using only the electrical fingerprint and making it autonomous against the cyberattack.

### 5.1. Data Set

We have collected 36 features from our EVSE architecture at a single timestamp. The features include all the electrical signals, control signals being used in the running EVSE system. The sampling time of data collection is set at $T_s = 10 \, \mu s$, thus the sampling frequency is $f_s = \frac{1}{T_s} = 10^5 \, samples/s$. Since the total simulation time is set to 15 seconds, the total number of samples belonging to a single attack class is $15 f_s = 15 x 10^5 = 1.5 \, M \, samples$. For four different attack classes ('Normal', 'PV Attack', 'BES Attack', 'EV Attack') we have dataset {x} of sample size of [1500000, 36, 4]. Table 3 presents the electrical fingerprint used for IDS.

**Table 3.** Datasets overview

| Components | Features | Total # |
|---|---|---|
| PV panel | $i_{pv}, v_{pv}, p_{pv}, i_{diode}, del_{in}, T, irradiance$ | 7 |
| MPPT Boost converter | $s_{boost}, \{i,v\}_{switch}, duty_{pv}, i_{bus}, v_{bus}, timest$ | 7 |
| BES | $soc, i, v, s_p, s_n, i_{ref}, duty_{bes}, v_{ref}, \{i,v\}_{switch}$ | 12 |
| EV | $soc, i, v, duty, i_{ref}, v_{ref}$ | 6 |
| Diodes | $\{i,v\}x2$ | 4 |

### 5.3. Stacked/Deep LSTM

The LSTM is the generalized state machine and de-facto RNN primarily designed for the regression or classification of the sequential data, i.e., sequence learning. The stacking of these individual LSTM cells into the hidden layers forms the Deep LSTM. A single LSTM unit is much more complex than a traditional neural unit. It has four gates: input gate, output gate, forget gate, and cell gate [35]. The LSTM cell takes input feature $x_t$ along with cell state $c_{t-1}$ and hidden state $h_{t-1}$ from previous LSTM units and outputs the current cell state $c_t$ and hidden state $h_t$.

The deep LSTM has stacked layers of multiple single LSTM cells with four gate parameters and an output parameter described below by eq 21-25 [36].

Input gate parameters: $\begin{pmatrix} W_{xi} \\ W_{hi} \end{pmatrix} \in \mathbb{R}^{(D+H)X H}, b_i \in \mathbb{R}^H$ (21)

Forget gate parameters: $\begin{pmatrix} W_{xf} \\ W_{hf} \end{pmatrix} \in \mathbb{R}^{(D+H)X H}, b_f \in \mathbb{R}^H$ (22)

Cell parameters: $\begin{pmatrix} W_{xc} \\ W_{hc} \end{pmatrix} \in \mathbb{R}^{(D+H)X H}, b_c \in \mathbb{R}^H$ (23)

Output gate parameters: $\begin{pmatrix} W_{xo} \\ W_{ho} \end{pmatrix} \in \mathbb{R}^{(D+H)X H}, b_o \in \mathbb{R}^H$ (24)

Network output layer parameters: $W_{hK} \in \mathbb{R}^{HX K}, b_K \in \mathbb{R}^K$ **(25)**

Where $W_{()}$ is the weight matrices, $b_{()}$ is the bias vectors, $D$ is the dimension of the input signal, $H$ is the number of LSTM units, and $K$ denotes the number of output classes. The respective outputs of the input, forget, cell, and output gates $\{i_t, f_t, \widetilde{c}_t, o_t\}$ in the forward pass can be written as follows in eqs. 26-30.

$$i_t = \sigma\left[\begin{pmatrix}W_{xi}\\W_{hi}\end{pmatrix}^T [x_t, h_{t-1}] + b_i\right] \quad (26)$$

$$f_t = \sigma\left[\begin{pmatrix}W_{xf}\\W_{hf}\end{pmatrix}^T [x_t, h_{t-1}] + b_f\right] \quad (27)$$

$$\widetilde{c}_t = \tanh\left[\begin{pmatrix}W_{xc}\\W_{hc}\end{pmatrix}^T [x_t, h_{t-1}] + b_c\right] \quad (28)$$

$$o_t = \sigma\left[\begin{pmatrix}W_{xo}\\W_{ho}\end{pmatrix}^T [x_t, h_{t-1}] + b_o\right] \quad (29)$$

$$h_t = o_t * \tanh(c_t) \quad (30)$$

Where $\sigma$ is a nonlinear function. The current LSTM outputs: cell and hidden states $\{c_t, h_t\}$ are passed to the next timestamps to iterate through the above equations. The probability vector $\{p_t\}_{k=1}^{k=K}$ for class $K$ can be computed by using SoftMax function as in eq. 31.

$$p_t = \text{SoftMax}(W_{hK}^T h_t + b_K) \quad (31)$$
$$\widehat{K} = \underset{k}{\text{argmax}} \; p_{tk} \quad (32)$$

The predicted class $\widehat{K}$ would be the one with the highest probability at timestamp t as shown below in eq. 32.

## 6. Simulation Procedure

This research work uses licensed NetSim Standard version 12.2 as a discrete event network simulator to simulate a 5G network through which an EVSE communicates with SCADA for command and control.

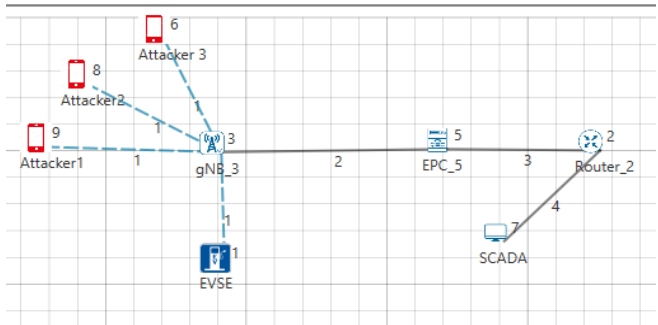

*Fig.9. Snapshot of Network Architecture with three attackers in NetSim.*

Each network component in a NetSim complies with Release 15/3GPP 38.xxx series and is flexible enough to specify user-defined longitude and latitude. The network components used are summarized below: Each smaller square box in Fig.9 represents the area with 10-meter longitude and 10-meter latitude.
**a) EVSE:** are equipped with Long term evolution new radio user equipment (LTE-NR-UE) device type with configurable mobility model set to be static. The application layer implements an open flow protocol and SDN controller, while the transport layer has a configurable TCP protocol. The network layer uses IPV4 protocol with a user-defined IP address, subnet mask, and default gateway. The physical layer of UE is equipped with configurable height, transmission power, and beamforming gain, respectively, set to 1.50 meters, 23 dBm, and 0 dB. The same configuration goes UE with rogue EVSE with spoofed IP.
**b) gNB:** Next-generation node B (gNB) is the 5G wireless base station that communicates with UE and 5G core network. The gNB height is 10 meters with a transmission power of 40 dBm. The wireless communication between UE and gNB is time division duplexing (TDD) with 15 kHz subcarrier spacing, the outdoor scenario of rural macro and channel characteristics with no path loss. The gNB is set to have a round-robin scheduling type with a UE measurement report of 120 milliseconds for the experimentation.
**c) EPC:** Evolved packet core (EPC) was introduced in 4G LTE and had a core network functionality.
**d) SCADA server:** This wireless node is the SCADA system that continuously gets logs of electrical signals from EVSE and sends control commands to manage the controllers at EVSE.

This setup is designed for the communication between the EVSE and remote control station at SCADA. The SCADA continuously monitors the operation of EVSE and issues the control commands through the 5G network. These experimental setups start with no attack scenario, i.e., a normal operating condition of 5G communication link, and the number of attackers increases from 0 to 15. The NetSim simulation time is set to run for 200 seconds to observe the delay. The physical layer setup in Simulink is already explained in section 2 with greater details.

## 7. Simulation results and discussion

In this work, the Syn flood attacks are triggered in the 5G enabled communication link between the SCADA node and the EVSE node. As evident from Table 4, The incremental change in delay with an increase in the number of attacks is more vehement. The throughput has been more consistent as opposed to latency. The throughput drops up to 20.94% compared to the base throughput. The latency has been increased significantly as the number of attackers increases from 0 to 15. The worst scenario of 509.476 ms (0.5 seconds) delay has been used to visualize the impact on the EVSE system.

**Table 4**. Network performance with increasing attack penetration

| Malicious nodes | dealy(millisec) | Throughput (Mbps) |
|---|---|---|
| 0 | 2.957 | 23.138 |
| 1 | 23.262 | 22.944 |
| 2 | 23.261 | 22.944 |
| 3 | 92.687 | 22.283 |
| 4 | 127.318 | 21.928 |
| 5 | 162.019 | 21.617 |
| 6 | 196.850 | 21.285 |
| 7 | 231.308 | 20.954 |
| 8 | 266.424 | 20.629 |
| 9 | 300.954 | 20.274 |
| 10 | 335.503 | 19.956 |
| 11 | 370.730 | 19.621 |
| 12 | 405.044 | 19.290 |
| 13 | 439.928 | 18.973 |

| 14 | 474.631 | 18.627 |
| 15 | 509.476 | 18.293 |

### 7.1. Impact analysis of FDI attacks

In this case, the simulation runs continuously for 15 seconds in Simulink, and attacks are launched at different controllers within the simulation time. The FDI attack analysis has been done in two scenarios to quantify the severity. i) Attacks are launched on different controllers at different times, and ii) Attacks are launched on different controllers simultaneously. The black plot in Fig. 10 represents the various electrical parameters during normal operation, while the red represents those parameters under the FDI attack.

i). FDI Attacks launched on different controllers at different times

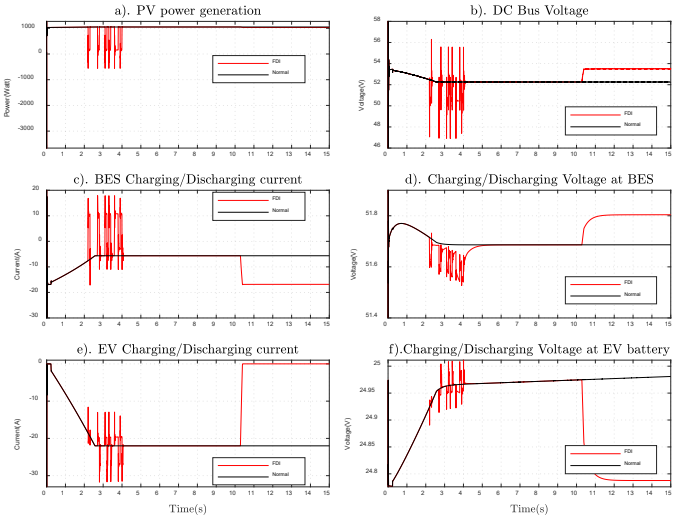

*Fig.10. Impacts of FDI attacks launched at PV controller from 2-4 seconds, BES controller from 6-8 seconds, and EV controller at 10 -12 seconds.*

PV controller attack: The duty cycle attack launched from 2 to 4 seconds at PV controller has caused severe ripples at PV power generation, excursing the power level to -500 Watt as shown in Fig.10a. The ripples at the PV controller contribute to the oscillating voltage at the DC bus bar by $\pm$ 5 V as in Fig.10b. Further ahead, the ripples pass on to the BES oscillates the current through the range of [-17, +17] A from the normal operating current of -5 A as in Fig.10 c. At the same time, the BES voltage goes down and oscillates, as in fig Fig.10d. Similarly, these low-frequency oscillations have a severe impact on the EV battery as the charging current has steep spikes and dips ranges from -31 A to -13 A within 105 ms as in Fig.10e. Fig.10f shows the slight oscillations at voltage during the attack. Therefore, the FDI attack at the PV controller can destabilize the entire EVSE ecosystem, i.e., PGU, ESU, and a plugged-in EV. However, the system gains its normal operating conditions as the attack goes off at t=4 seconds.

BES controller attack: The FDI attack at the BES controller does not seem to have any significant impact as the finely tuned PI controller saturates the incoming fluctuation from t=6-8 seconds in Fig.10e.

EV controller attack: The EV controller attack starts at t=10 seconds and lasts for 2 seconds, as in Fig.10. This impact is irreversible and does not affect PGU, except it increases the bus voltage by 2V (Fig.10b). The attack has risen sharply the constant charging current at BES by 11 A (Fig.10c) with a slight DC shift of 117 mv (Fig.10.d). The attack at EV has shifted constant charging battery current at -21 A to 0.268 µA (Fig.10e) assisted by slight voltage drop (Fig.10f), i.e., forces charging EV to stop charging completely.

ii). FDI Attacks launched on different controllers simultaneously:

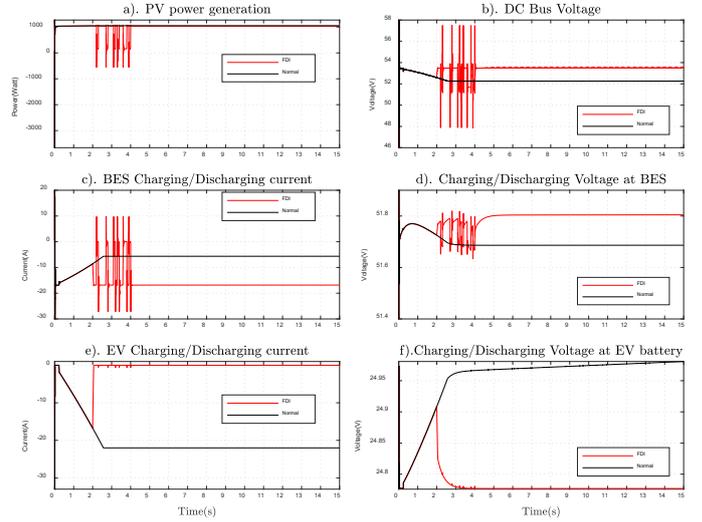

*Fig.11. Impacts of FDI attacks launched at all controllers simultaneously from 2-4 seconds.*

In this case, all the controllers are attacked simultaneously from 2 to 4 seconds as in Fig.11. The integrated attacks resulted in reversible low-frequency oscillations throughout the attack and irreversible DC shift. The magnitude and frequency of oscillation at power generation remain the same Fig.11a. The DC bus voltage has spikes of equal magnitude and shifted up by 2V, as shown in Fig.11b as opposed to the standalone attack at PV control.

The variation of BES current ranges from [ -25 A 8.5 A] from normal constant charging of -5.6 A. It signifies the frequent charging and discharging of BES within a short time span. The peaks are even, and the constant charging current has been increased by 300% even after the attack that never comes back to normal mode, as shown in Fig.11c. The BES voltage oscillations are the same as the standalone attack at PV except for an irreversible DC shift of 117 mV, as shown in Fig.11d.

The EV charging current has been found to be dropped to zero permanently from -22 A with slight oscillation around 0-2 seconds, as shown in Fig.11e. That means the EV has stopped charging. The minimal temporary oscillation has been observed in EV battery voltage with a slight irreversible drop shown in Fig.11f.

### 7.2. Impact analysis of DDoS attack

Under similar simulation setups as the FDI attack, different delays resulted from a cyberattack on 5G in NetSim has been tested in our EVSE system. Likewise, the DDoS attack is carried out in all three control components simultaneously and at different times. Fig. 12 presents the impact of DDoS attacks on the physical system caused by the 500 ms delay of the 5G communication system.

The attacker launches the DDoS attack on the PV controller at t=2 seconds and lasts for 500 ms. This attack causes high-frequency oscillations at power signal that fluctuates between -3.6 kW to 1.277 kW, while the PV generation system was designed to deliver 1.065 kW, as shown in Fig.12a. The negative power means the PV is drawing the power from the EVSE. Similarly, these high-frequency oscillation causes momentary voltage swing and drop of 2V at DC bus as shown in Fig.12b. At BES, the current surges by 17 A as shown in Fig.12c, and voltage drops by 200 mV as shown in Fig.12d.

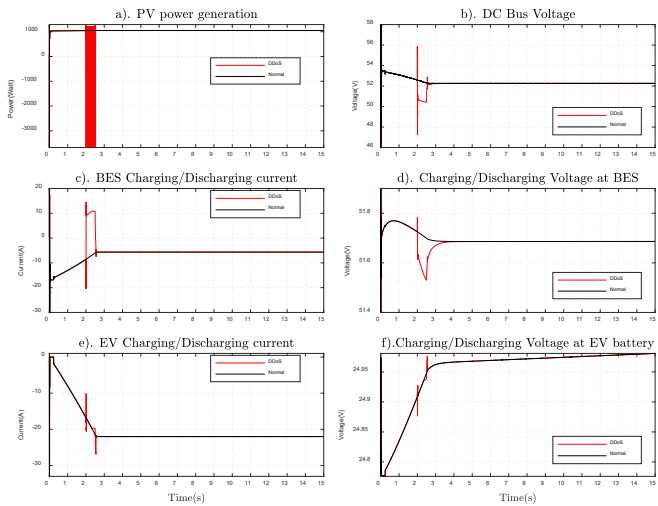

*Fig.12.* Impacts of DDoS attacks launched at PV controller from 2-2.5 seconds, BES controller from 6-6.5 seconds, and EV controller at 10 -10.5 seconds.

Similarly, spikes of -10 A and -26 A were observed at the beginning and end of the attack respectively in charging current as shown in Fig.13e and is accompanied by small complementary voltage surges in EV battery as shown in Fig.13f. As soon as the attack stops, the system completely comes back.

The attacks at BES control and EV controller have no impact on system response. This is because the PI controller has fixed upper and lower saturation thresholds that do not let the manipulated $V_{ref}$ signal to produce zero control signal, though the signal is completely lost throughout the attack. Our experiment suggests that the improperly tuned PI controller with no saturation thresholds is found to be exploited by the DDoS attack.

### 7.3. IDS performance analysis

The deep learning algorithms are created in Python 3.7.4 in the Jupyter lab (version 1.1.4) under the free and open-source Anaconda distribution. Intel® Core™ i7-9750 @ 2.60 GHz processor with 16.00 GB RAM and 64-bit Windows 10 OS is used.

Fig.13 represents the accuracy and loss progression during the training and validation of the model. The proposed model achieves more than 99.9999% accuracy within the fifth epochs for both training and validation data. The loss is diminished to 6.11e-06 within fifth epochs for both training and validation datasets. This signifies the smoother progression during the training, and our model is ready to classify the previously unseen datasets of the attack.

The classifier's testing or generalization performance can be better presented with the confusion matrix, as shown in Fig.14. The distribution of test data among different kinds of attacks is almost equiproportional, i.e., $\cong$ 25% from each class. The y-axis represents the predicted class, and the x-axis represents the actual class. The numbers and % in each cell represent the number of samples and % of samples belonging to that cell. For instance, for the first cell, among 449,234 normal samples fetched to our model. It correctly predicts 449,232 sample instances, leaving only one sample misclassified as PV attack and another sample misclassified as BES Attack, which is almost 100% classification accuracy. Each row in the last column represents the total number of samples fetched to the classifier. Each column of the last row sums up the total number of classified samples belonging to different classes. Our classifier detects PV Attack, BES Attack, EV Attack with 100% accuracy while detecting normal data with 99.99999% accuracy.

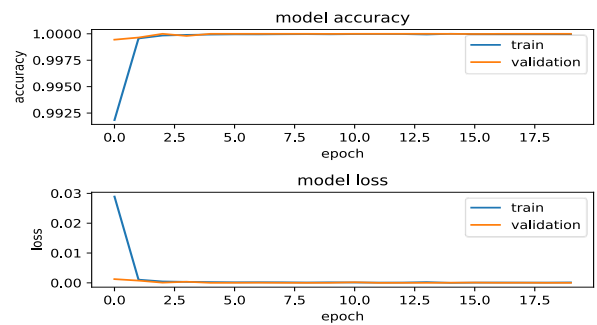

*Fig.13.* Accuracy and loss during Training and Validation progression

*Fig. 14* Confusion Matrix for assessing model performance

As per table 5 below, the proposed classifier has almost 100% precision implying the model's repeatability; 100% recall implying capability to correctly classify attack among different classes of attack; and 100 % f1-score indicating superior sensitivity and separability of the model.

**Table 5.** Classification Metrics

| Attack Class | Precision | Recall | F1-score | support |
|---|---|---|---|---|
| Normal | 1.00 | 1.00 | 1.00 | 449232 |
| PV Attack | 1.00 | 1.00 | 1.00 | 450398 |

| BES Attack | 1.00 | 1.00 | 1.00 | 449922 |
| EV Attack | 1.00 | 1.00 | 1.00 | 450450 |

All the proposed IDS performance metrics are found to be superior compared to our past IDS [11].

The most important capability of our proposed IDS is that it can detect the attempt of attack that bypasses the cyber layer and is not noticed in the monitoring station. For instance, the FDI attack at BES from 6-8 seconds, as shown in Fig 11, does not seem to change any electrical parameters, though the attack is there. But our IDS can detect it with 100% precision, recall, and f1-score does not even misclassify a single sample. It's because of including the control signals as features.

## 8. Conclusion

This work analyzes the cybersecurity issues in the 5G enabled EVSE system with a deep learning-based attack detection method. The 5G enabled standalone off-the-grid PV-powered EVSE architecture has been built and simulated to charge the PEV. The FDI and DDoS attacks have been successfully launched and simulated in the 5G-powered EVSE communicating with remote SCADA, and consequent impact analysis has been presented. Finally, the stacked LSTM based local IDS has been developed and tested on the proposed EVSE solely based on electrical fingerprints. The following conclusions can be drawn from the study:

1) The low-frequency FDI attack on the PV controller's duty cycle produces ripples and impacts all the subsequent components throughout the attack.
2) The FDI attack at BES has no visible impact since the PI controller operates around the saturation region to cope with the attack.
3) The FDI attack at the EV controller has resulted in an irreversible DC shift in operating current and voltage.
4) The simultaneous FDI attacks at all controllers have the integrated impact of points 1, 2, and 3.
5) The DDoS attack at the PV controller has caused high-frequency oscillations at PV power generation and high magnitude spikes and dips to subsequent Bus, BES, and EV controllers throughout the attack.
6) The DDoS attack at EV and BES has no impact due to PI controller saturation.
7) The proposed Stacked LSTM based IDS has been tested effective and can detect the bypassed cyber attack as well as unnoticed stealthy cyberattack attempts.

This novel application could safeguard the EVCS and its stakeholders from possible cyber threats. Adding new kinds of attack data in training, the proposed model could easily scale up to detect more diverse attacks. It ensures the scalability and interoperability of our model.

Our future research direction incldes building the independent, self-reliant, agile, trustworthy and resilient EVSE integrated with transportation and grid system. Moreover, the focus should be on developing the control algorithms to minimize the attack impact on the EVSE system.

## 9. Acknowledgments

The authors would like to thank the University of Memphis for providing Carnegie R1 doctoral fellowship and all required resources to complete this research.

## 10. References

[1] K. Harnett, B. Harris, D. Chin, and G. Watson, "DOE/DHS/DOT Volpe Technical Meeting on Electric Vehicle and Charging Station Cybersecurity Report," p. 44.

[2] "Securing Vehicle Charging Infrastructure APR," *Cyber Secur.*, p. 6, 2019.

[3] S. Vitturi, C. Zunino, and T. Sauter, "Industrial Communication Systems and Their Future Challenges: Next-Generation Ethernet, IIoT, and 5G," *Proc. IEEE*, vol. 107, no. 6, pp. 944–961, Jun. 2019, doi: 10.1109/JPROC.2019.2913443.

[4] P. Popovski, K. F. Trillingsgaard, O. Simeone, and G. Durisi, "5G Wireless Network Slicing for eMBB, URLLC, and mMTC: A Communication-Theoretic View," *IEEE Access*, vol. 6, pp. 55765–55779, 2018, doi: 10.1109/ACCESS.2018.2872781.

[5] G. Naik, B. Choudhury, and J. Park, "IEEE 802.11bd 5G NR V2X: Evolution of Radio Access Technologies for V2X Communications," *IEEE Access*, vol. 7, pp. 70169–70184, 2019, doi: 10.1109/ACCESS.2019.2919489.

[6] M. Polese et al., "Integrated Access and Backhaul in 5G mmWave Networks: Potential and Challenges," *IEEE Commun. Mag.*, vol. 58, no. 3, pp. 62–68, Mar. 2020, doi: 10.1109/MCOM.001.1900346.

[7] S. Lagen et al., "New Radio Beam-Based Access to Unlicensed Spectrum: Design Challenges and Solutions," *IEEE Commun. Surv. Tutor.*, vol. 22, no. 1, pp. 8–37, Firstquarter 2020, doi: 10.1109/COMST.2019.2949145.

[8] "Level 1 and Level 2 Electric Vehicle Service Equipment (EVSE) Reference Design," p. 36, 2016.

[9] S. Hu, X. Chen, W. Ni, X. Wang, and E. Hossain, "Modeling and Analysis of Energy Harvesting and Smart Grid-Powered Wireless Communication Networks: A Contemporary Survey," *IEEE Trans. Green Commun. Netw.*, vol. 4, no. 2, pp. 461–496, Jun. 2020, doi: 10.1109/TGCN.2020.2988270.

[10] R. Heartfield, G. Loukas, and D. Gan, "You Are Probably Not the Weakest Link: Towards Practical Prediction of Susceptibility to Semantic Social Engineering Attacks," *IEEE Access*, vol. 4, pp. 6910–6928, 2016, doi: 10.1109/ACCESS.2016.2616285.

[11] M. Basnet and M. H. Ali, "Deep Learning-based Intrusion Detection System for Electric Vehicle Charging Station," in *2020 2nd International Conference on Smart Power Internet Energy Systems (SPIES)*, Sep. 2020, pp. 408–413, doi: 10.1109/SPIES48661.2020.9243152.

[12] S. Mousavian, M. Erol-Kantarci, L. Wu, and T. Ortmeyer, "A Risk-Based Optimization Model for Electric Vehicle Infrastructure Response to Cyber Attacks," *IEEE Trans. Smart Grid*, vol. 9, no. 6, pp.

[13] D. Reeh, F. C. Tapia, Y. Chung, B. Khaki, C. Chu, and R. Gadh, "Vulnerability Analysis and Risk Assessment of EV Charging System under Cyber-Physical Threats," in *2019 IEEE Transportation Electrification Conference and Expo (ITEC)*, Jun. 2019, pp. 1–6, doi: 10.1109/ITEC.2019.8790593.

[14] A. Huseinovic, S. Mrdovic, K. Bicakci, and S. Uludag, "A Taxonomy of the Emerging Denial-of-Service Attacks in the Smart Grid and Countermeasures," in *2018 26th Telecommunications Forum (TELFOR)*, Belgrade, Nov. 2018, pp. 1–4, doi: 10.1109/TELFOR.2018.8611847.

[15] H.-J. Liao, C.-H. Richard Lin, Y.-C. Lin, and K.-Y. Tung, "Intrusion detection system: A comprehensive review," *J. Netw. Comput. Appl.*, vol. 36, no. 1, pp. 16–24, Jan. 2013, doi: 10.1016/j.jnca.2012.09.004.

[16] A. Khraisat, I. Gondal, P. Vamplew, and J. Kamruzzaman, "Survey of intrusion detection systems: techniques, datasets and challenges," *Cybersecurity*, vol. 2, no. 1, p. 20, Dec. 2019, doi: 10.1186/s42400-019-0038-7.

[17] S. Acharya, Y. Dvorkin, H. Pandžić, and R. Karri, "Cybersecurity of Smart Electric Vehicle Charging: A Power Grid Perspective," *IEEE Access*, vol. 8, pp. 214434–214453, 2020, doi: 10.1109/ACCESS.2020.3041074.

[18] R. Gottumukkala, R. Merchant, A. Tauzin, K. Leon, A. Roche, and P. Darby, "Cyber-physical System Security of Vehicle Charging Stations," in *2019 IEEE Green Technologies Conference(GreenTech)*, Lafayette, LA, USA, Apr. 2019, pp. 1–5, doi: 10.1109/GreenTech.2019.8767141.

[19] D. Niyato, D. T. Hoang, P. Wang, and Z. Han, "Cyber Insurance for Plug-In Electric Vehicle Charging in Vehicle-to-Grid Systems," *IEEE Netw.*, vol. 31, no. 2, pp. 38–46, Mar. 2017, doi: 10.1109/MNET.2017.1600321NM.

[20] J. Antoun, M. E. Kabir, B. Moussa, R. Atallah, and C. Assi, "A Detailed Security Assessment of the EV Charging Ecosystem," *IEEE Netw.*, vol. 34, no. 3, pp. 200–207, May 2020, doi: 10.1109/MNET.001.1900348.

[21] Y. Shen, W. Fang, F. Ye, and M. Kadoch, "EV Charging Behavior Analysis Using Hybrid Intelligence for 5G Smart Grid," *Electronics*, vol. 9, no. 1, Art. no. 1, Jan. 2020, doi: 10.3390/electronics9010080.

[22] N. Femia, G. Petrone, G. Spagnuolo, and M. Vitelli, "A Technique for Improving P O MPPT Performances of Double-Stage Grid-Connected Photovoltaic Systems," *IEEE Trans. Ind. Electron.*, vol. 56, no. 11, pp. 4473–4482, Nov. 2009, doi: 10.1109/TIE.2009.2029589.

[23] O. Rabiaa, B. H. Mouna, S. Lassaad, F. Aymen, and A. Aicha, "Cascade Control Loop of DC-DC Boost Converter Using PI Controller," in *2018 International Symposium on Advanced Electrical and Communication Technologies (ISAECT)*, Nov. 2018, pp. 1–5, doi: 10.1109/ISAECT.2018.8618859.

[24] N. Femia, G. Petrone, G. Spagnuolo, and M. Vitelli, "Optimizing duty-cycle perturbation of P O MPPT technique," in *2004 IEEE 35th Annual Power Electronics Specialists Conference (IEEE Cat. No.04CH37551)*, Jun. 2004, vol. 3, pp. 1939-1944 Vol.3, doi: 10.1109/PESC.2004.1355414.

[25] K. J. Sauer and T. Roessler, "Systematic approaches to ensure correct representation of measured multi-irradiance module performance in PV system energy production forecasting software programs," in *2012 38th IEEE Photovoltaic Specialists Conference*, Jun. 2012, pp. 000703–000709, doi: 10.1109/PVSC.2012.6317706.

[26] V. Kumar, V. R. Teja, M. Singh, and S. Mishra, "PV Based Off-Grid Charging Station for Electric Vehicle," *IFAC-Pap.*, vol. 52, no. 4, pp. 276–281, Jan. 2019, doi: 10.1016/j.ifacol.2019.08.211.

[27] A. Ghosh, A. Maeder, M. Baker, and D. Chandramouli, "5G Evolution: A View on 5G Cellular Technology Beyond 3GPP Release 15," *IEEE Access*, vol. 7, pp. 127639–127651, 2019, doi: 10.1109/ACCESS.2019.2939938.

[28] "Updated ENISA 5G Threat Landscape Report to Enhance 5G Security." https://www.enisa.europa.eu/news/enisa-news/updated-enisa-5g-threat-landscape-report-to-enhance-5g-security (accessed Dec. 22, 2020).

[29] P. T. Krein and M. A. Fasugba, "Vehicle-to-grid power system services with electric and plug-in vehicles based on flexibility in unidirectional charging," vol. 1, no. 1, p. 11, 2017.

[30] T. Liu and T. Shu, "Adversarial FDI Attack against AC State Estimation with ANN," *ArXiv190611328 Cs Stat*, Jun. 2019, Accessed: Apr. 07, 2021. [Online]. Available: http://arxiv.org/abs/1906.11328.

[31] P.-Y. Chen, S. Yang, J. A. McCann, J. Lin, and X. Yang, "Detection of false data injection attacks in smart-grid systems," *IEEE Commun. Mag.*, vol. 53, no. 2, pp. 206–213, Feb. 2015, doi: 10.1109/MCOM.2015.7045410.

[32] M. Bogdanoski, T. Suminoski, and A. Risteski, "Analysis of the SYN Flood DoS Attack," *Int. J. Comput. Netw. Inf. Secur. IJCNIS*, vol. 5, no. 8, Art. no. 8, Jun. 2013.

[33] A. Ingle and M. Awade, "Intrusion detection for TCP-SYNC Flood attack," *Int. J. Adv. Res. Comput. Sci.*, vol. 4, no. 5, May 2013, Accessed: Jan. 05, 2021. [Online]. Available: https://search.proquest.com/docview/1443755249/abstract/F9C268C2959B4E2APQ/1.

[34] N. R. Projects, *NetSim-TETCOS/DOS_Attack_in_5G_v12.1*. 2020.

[35] Y. Yan, L. Qi, J. Wang, Y. Lin, and L. Chen, "A Network Intrusion Detection Method Based on Stacked Autoencoder and LSTM," in *ICC 2020 - 2020 IEEE International Conference on Communications (ICC)*, Jun. 2020, pp. 1–6, doi: 10.1109/ICC40277.2020.9149384.

[36] Y. Guan and T. Plötz, "Ensembles of Deep LSTM Learners for Activity Recognition using Wearables," *Proc. ACM Interact. Mob. Wearable Ubiquitous Technol.*, vol. 1, no. 2, p. 11:1-11:28, Jun. 2017, doi: 10.1145/3090076.